\documentclass[proof]{WileyASNA-v2}

\articletype{Original Article}

\received{X X 2021}
\revised{X X 2021}
\accepted{X X 2021}

\usepackage[T1]{fontenc}
\usepackage[utf8]{inputenc}
\usepackage[american]{babel}

\begin{document}

\title{Bright Spectroscopic Binaries: II. A study of five systems with orbital periods of $P\lesssim500$~days}

\author[1]{Dennis Jack*}
\author[1]{Missael Alejandro Hern\'andez Huerta}
\author[1]{Faiber Danilo Rosas-Portilla}
\author[1,2]{Klaus-Peter Schr\"oder}

\authormark{D. Jack et al.}

\address[1]{\orgdiv{Departamento de Astronom\'\i{}a}, \orgname{Universidad de Guanajuato}, \orgaddress{\state{Guanajuato}, \country{Mexico}}}
\address[2]{\orgdiv{Sterrewacht Leiden}, \orgname{Universiteit Leiden}, \orgaddress{\state{Leiden}, \country{Netherlands}}}

\corres{*Dennis Jack, \email{dennis.jack@ugto.mx}}

\presentaddress{Departamento de Astronom\'\i{}a, Universidad de Guanajuato, Callej\'on de Jalisco S/N, 36023 Guanajuato, GTO, Mexico}

\abstract{We present a detailed analysis of five bright spectroscopic binary systems (HD~18665, HD~27131, HD~171852, HD~215550, HD~217427) 
that have orbital periods of $P \lesssim 500$~days. 
We determined the complete set of orbital parameters using the toolkit RadVel by analyzing the observed radial velocity curves. 
To study the properties of the five systems, we also analyzed the intermediate resolution spectra ($R\approx 20,000$) 
observed with the TIGRE telescope
and determined the stellar parameters of the primary stars using the toolkit iSpec. 
With Gaia Early Data Release~3 parallaxes, a correction for interstellar extinction using the 3D dust map,
and bolometric corrections, we placed the stars in the Hertzsprung-Russell diagram
and compared the positions with stellar evolution tracks calculated with the Eggleton code to determine the masses and ages of the primary stars. 
They have all evolved to the giant phase.
Finally, we were able to determine the masses of the secondary stars 
and to estimate the orbital inclinations $i$ of the binary systems. }


\fundingInfo{DFG-CONACYT}

\maketitle

\footnotetext{\textbf{Abbreviations:} RV, radial velocity;}

%
%
\section{Introduction}

Multiple stellar systems are a natural outcome of the star formation process \citep{tohline02,binreview},
and a binary fraction of 50\%  in the solar neighborhood has been estimated by \citet{raghavan10}.
\citet{fisher05} showed that well-separated
binaries like in this paper form a large fraction of the stellar galactic content.
Binary or multiple stars are important objects, 
because they offer the opportunity to determine the masses of the components directly.
This is impossible for single stellar objects.
The stellar mass is an important parameter for the understanding of the evolution of stars.

There exist several detection methods for binary systems.
One important method is the detection of periodic variations in the radial velocity (RV) by spectroscopic analysis of the stars.
In fact, the first exoplanet orbiting a main-sequence star was also discovered with this method \citep{51peg}.
Detecting and studying spectroscopic binary systems requires an intermediate to high spectral resolution, 
but the signal-to-noise ratio (S/N) can still be relatively low. A wide spectral range is also
helpful for a more precise determination of the RV using a large number of spectral lines. 
Several observations at different epochs are necessary
to detect the RV changes and to confirm the binarity of the star system. 
Usually, only the primary star is visible in the spectrum (SB1), but if the two stars have similar
luminosities, spectral lines of the secondary star will appear in the observed spectrum (SB2).
It is possible to determine the orbital parameters 
using many observations at different epochs obtaining a well covered RV curve.
This allows conclusions about the invisible secondary star in case of an SB1 system, which
are the objects we were studying in this publication.

In our first publication of this series \citep{paper1} (hereafter Paper~I), we presented details of our observation campaign
with the TIGRE telescope and reported the discovery of 19 bright binary star systems ($m_V < 7.66$~mag).
Because they have very different orbital periods, we presented a detailed analysis of only five binary systems 
with short orbital periods of less than one year in Paper~I.
In the present publication, we show the results of further five bright binary systems 
with orbital periods of $P \lesssim 500$~days. 
The remaining bright binary stars have even longer periods,
and we will continue with our observation campaign to determine their orbital parameters and will present the results in following
publications.

For this work, we improved several aspects of our methods for the analysis of the RV curves and the determination of stellar parameters.
One important improvement is the use of the recently published Gaia EDR3 catalogue \citep{gaiaedr3,gaiaedr31}
of the ESA Gaia mission \citep{gaiamission}, because it contains improved parallaxes for our stars.
To obtain more precise luminosities, we now correct the effects of interstellar extinction using the 3D dust mapping. 
Furthermore, we improved our method for the determination of the stellar parameters, like effective temperature and surface gravity.

%
%
\section{Binary Orbits}\label{sec:orb}

\subsection{Radial velocity curves}

The stars were observed with the 1.2~m robotic telescope TIGRE \citep{schmitt14} situated near Guanajuato, Central Mexico.
It is equipped with the HEROS spectrograph with an intermediate resolution of $R\approx 20,000$.
We obtained a time series of optical spectra in the wavelength range from about $3800$ to $8800$~\AA,
with a small gap of about 120~\AA\ around 5800~\AA. 
The exposure time of each spectrum was one minute. 
The RVs were determined with the procedure described in Paper~I and in \citet{mittag18}.
All individual RV measurements can be found in Table~\ref{tab:rv_1}
and are also electronically available as supporting
online material at \emph{link to Wiley page here} as well as at \url{http://www.astro.ugto.mx/~dennis/binaries/}.

Note that HD~27131 has already been identified as a spectroscopic binary by \citet{houk99}.
They state that the spectrum is a composition of a K0 giant and an A3 star.
We confirm that HD~27131 is in fact an SB2 star.
However, no further observations, detailed RV measurements, or orbital parameters had been reported, so far.

\subsection{Orbital parameters}

We used the Radial Velocity Modeling Toolkit RadVel \citep{radvel} (version 1.4.6)
to analyze the RV curves and to determine the orbital parameters of the binaries.
These parameters were the orbital period $P$, the time of inferior conjunction $T_c$,
eccentricity $e$, semi-amplitude $K$, the argument of the periapsis
of the star's orbit $\omega$ and the RV of the system $v_\mathrm{rad}$.
After inserting initial guesses for the six orbital parameters, 
the set of equations for the Keplerian orbit was solved by
an iterative method to determine the best fit for
the observed RV curve.
To obtain an estimation of the uncertainties, we used the Markov-Chain Monte Carlo (MCMC)
package of \citet{foreman2013} included in RadVel.

\begin{table*}[t]%
\centering
\caption{The orbital parameters of five spectroscopic binary systems determined with RadVel.}
\tabcolsep=0pt%
\begin{tabular*}{500pt}{@{\extracolsep\fill}lcccccc@{\extracolsep\fill}}
\toprule
\textbf{Star} & $P$ (days) & $T_c$ (JD) & $e$ & $K$ (km~s$^{-1}$) & $\omega$ (rad) & $v_\mathrm{rad}$ (km~s$^{-1}$) \\ 
\midrule
HD 18665 &  $506.0  \pm 1.5$  & $8308.5  \pm 3.3$  & $0.482  \pm 0.013 $ & $ 8.00  \pm 0.14$  & $5.611  \pm 0.029$  & $-5.052  \pm 0.072$ \\
HD 27131 &  $464.1  \pm 1.6$  & $8229.9  \pm 3.2$  & $0.1215 \pm 0.0092$ & $20.96  \pm 0.24$  & $5.315  \pm 0.077$  & $ 22.43  \pm 0.21$  \\
HD 171852 & $348.58 \pm 0.16$ & $8275.93 \pm 0.57$ & $0.2651 \pm 0.0032$ & $14.231 \pm 0.058$ & $4.89   \pm 0.01$  & $-27.515  \pm 0.034$ \\
HD 215550 & $436.76 \pm 0.24$ & $8716.86 \pm 0.23$ & $0.3619 \pm 0.0023$ & $18.243 \pm 0.045$ & $1.5307 \pm 0.0081$ & $-0.182  \pm 0.036$ \\
HD 217427 & $416.5  \pm 1.2$  & $8369.7  \pm 3.3$  & $0.415  \pm 0.013 $ & $9.13   \pm 0.17$  & $4.501  \pm 0.024$  & $-25.01  \pm 0.08$  \\
\bottomrule
\end{tabular*}
\label{tab:orbits}
\end{table*}

In Table~\ref{tab:orbits}, we present the results of the RadVel run
including the errors that we obtained with the MCMC run. 
One star has an orbital period of less than one year, but was not presented in Paper~I.
The orbits of the stars have a notable eccentricity $e$ that allowed a determination of their argument of the periapsis $\omega$. 
The semi-amplitudes $K$ of the orbits range from 8 to almost 21~km~s$^{-1}$.

\begin{figure*}[t]
    \centerline{\includegraphics[width=0.9\textwidth]{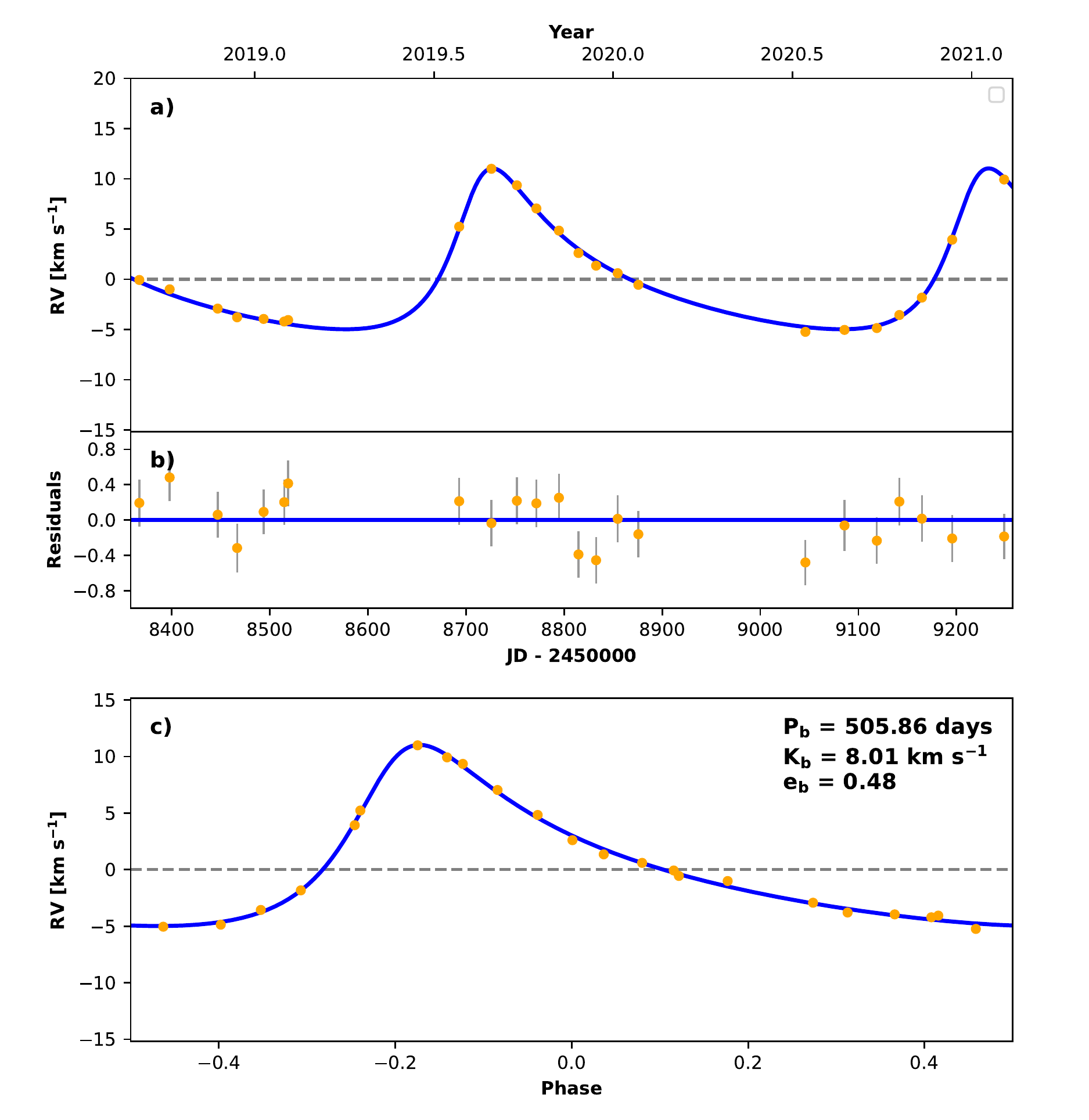}}
        \caption{RV curve of HD~18655: a) the complete measurements (circles) and RadVel fit (solid line),
        b) the residuals and c) the phase-folded RV curve.\label{fig:HD18665}}
\end{figure*}

In Fig.~\ref{fig:HD18665}, we plotted the observed RV curve and the RadVel fit of HD~18665.
\begin{figure*}[t]
    \centerline{\includegraphics[width=0.9\textwidth]{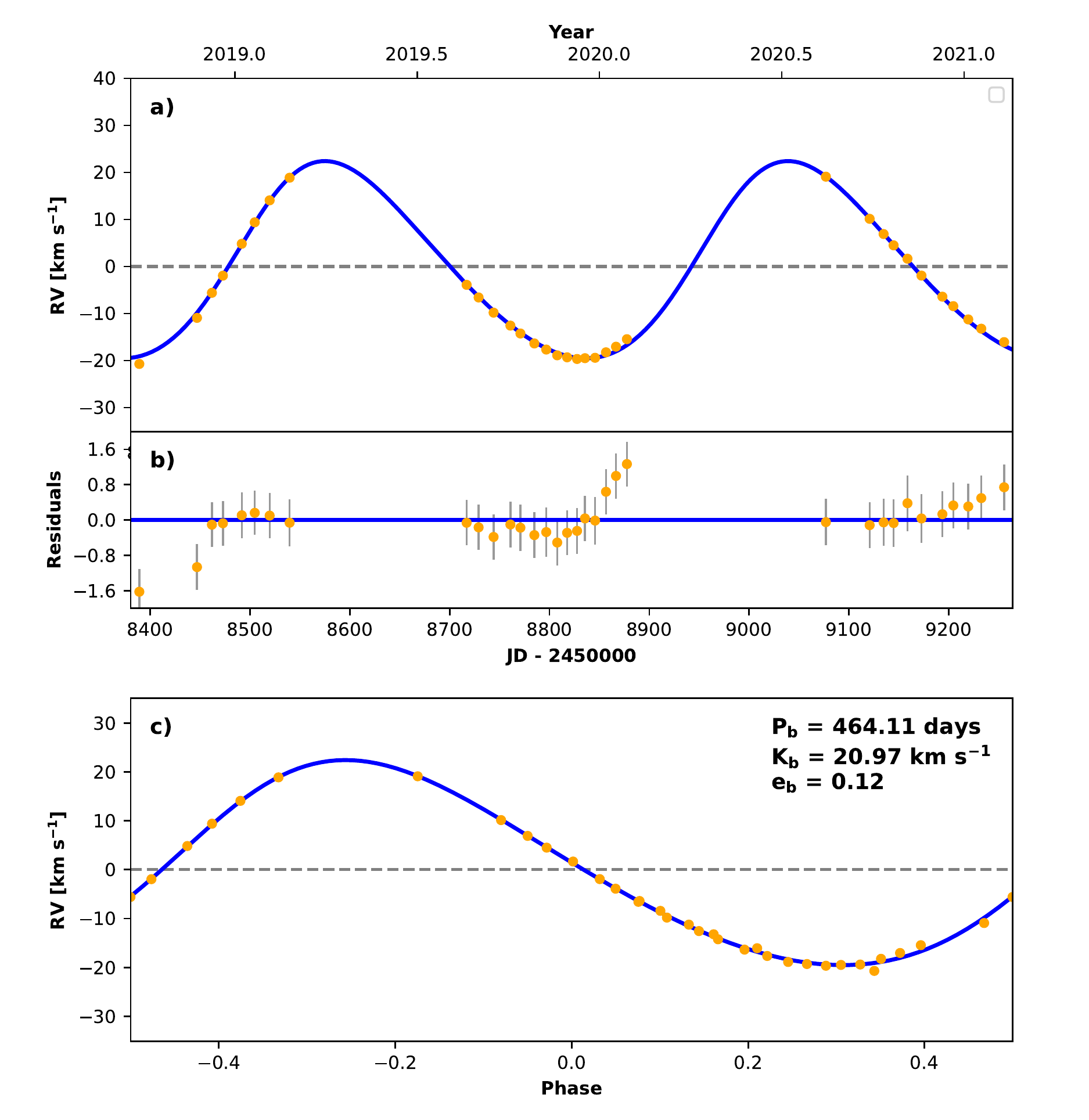}}
        \caption{RV curve of HD~27131: a) the complete measurements (circles) and RadVel fit (solid line),
        b) the residuals and c) the phase-folded RV curve.\label{fig:HD27131}}
\end{figure*}
The star HD~27131 has the largest semi-amplitude ($K=20.96$~km~s$^{-1}$) of all five stars and a low eccentricity
as can be seen in the RV curve in Fig.~\ref{fig:HD27131}.
\begin{figure*}[t]
    \centerline{\includegraphics[width=0.9\textwidth]{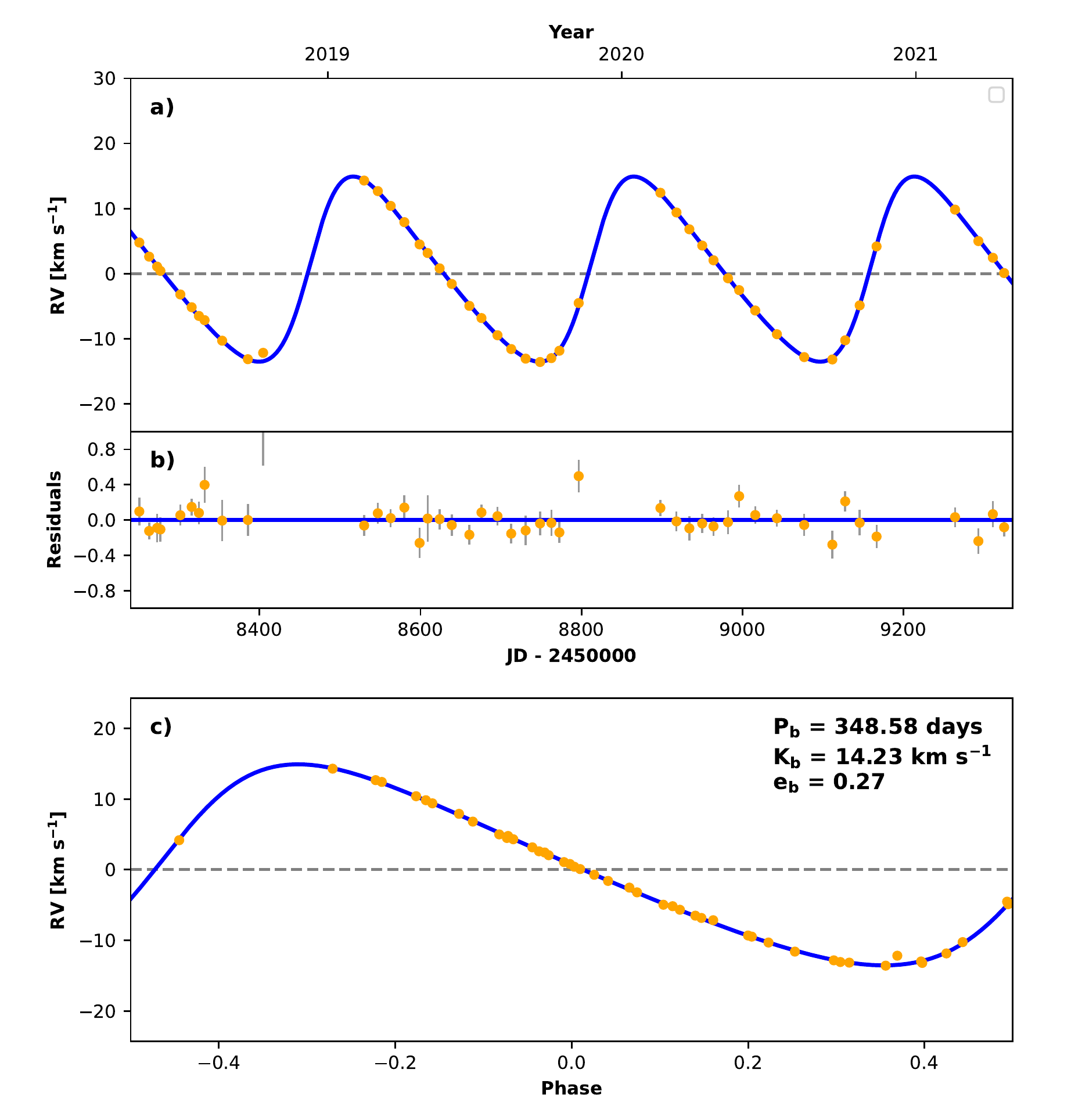}}
        \caption{RV curve of HD~171852: a) the complete measurements (circles) and RadVel fit (solid line),
        b) the residuals and c) the phase-folded RV curve.\label{fig:HD171852}}
\end{figure*}
In Fig.~\ref{fig:HD171852}, we present the complete RV measurements of the star HD~171852.
The orbital period is very close to one year, and the RV measurements cover more then two orbits.
\begin{figure*}[t]
    \centerline{\includegraphics[width=0.9\textwidth]{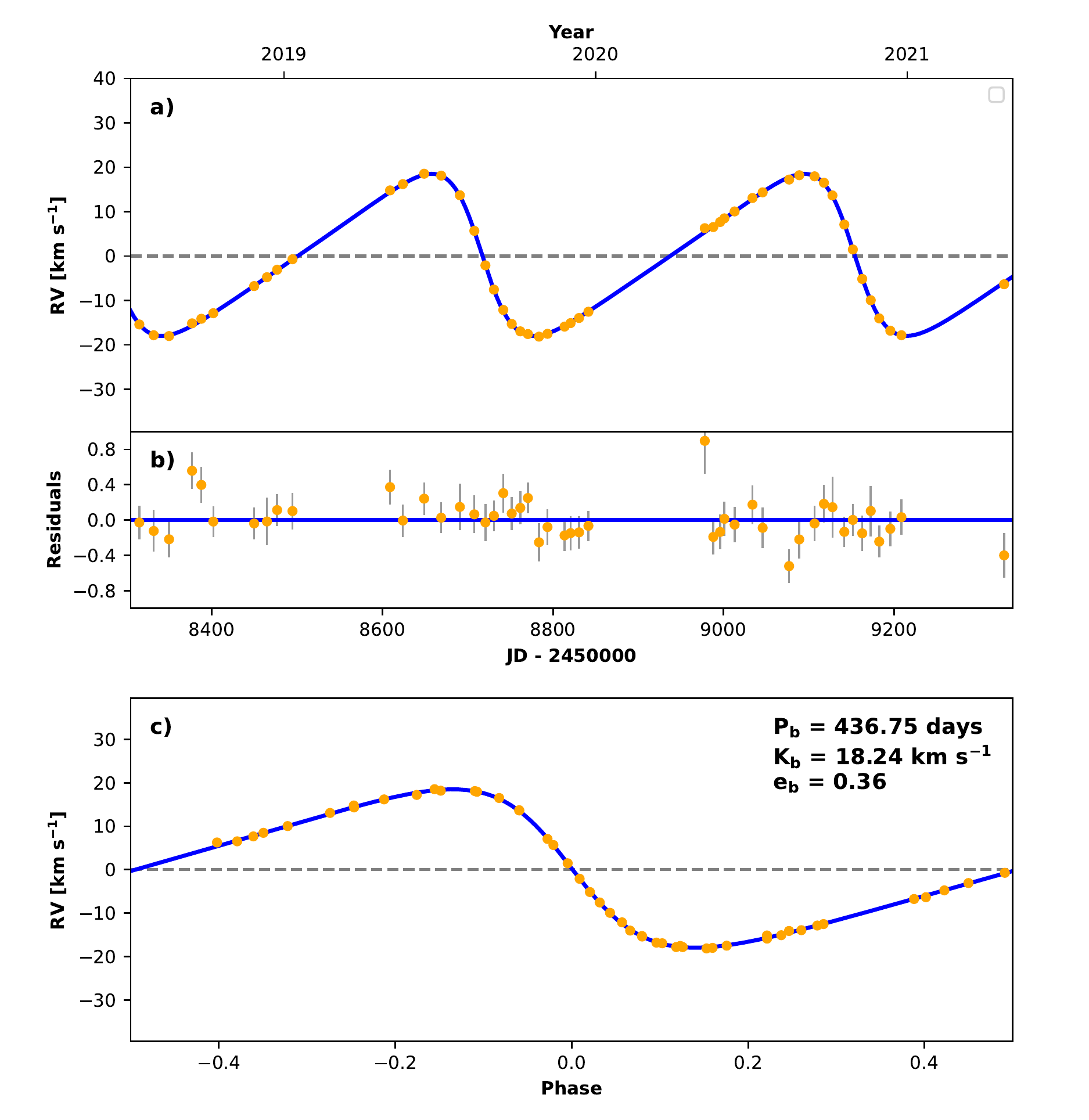}}
        \caption{RV curve of HD~215550: a) the complete measurements (circles) and RadVel fit (solid line),
        b) the residuals and c) the phase-folded RV curve.\label{fig:HD215550}}
\end{figure*}
Figure~\ref{fig:HD215550} demonstrate the RV curve of HD~215550.
\begin{figure*}[t]
    \centerline{\includegraphics[width=0.9\textwidth]{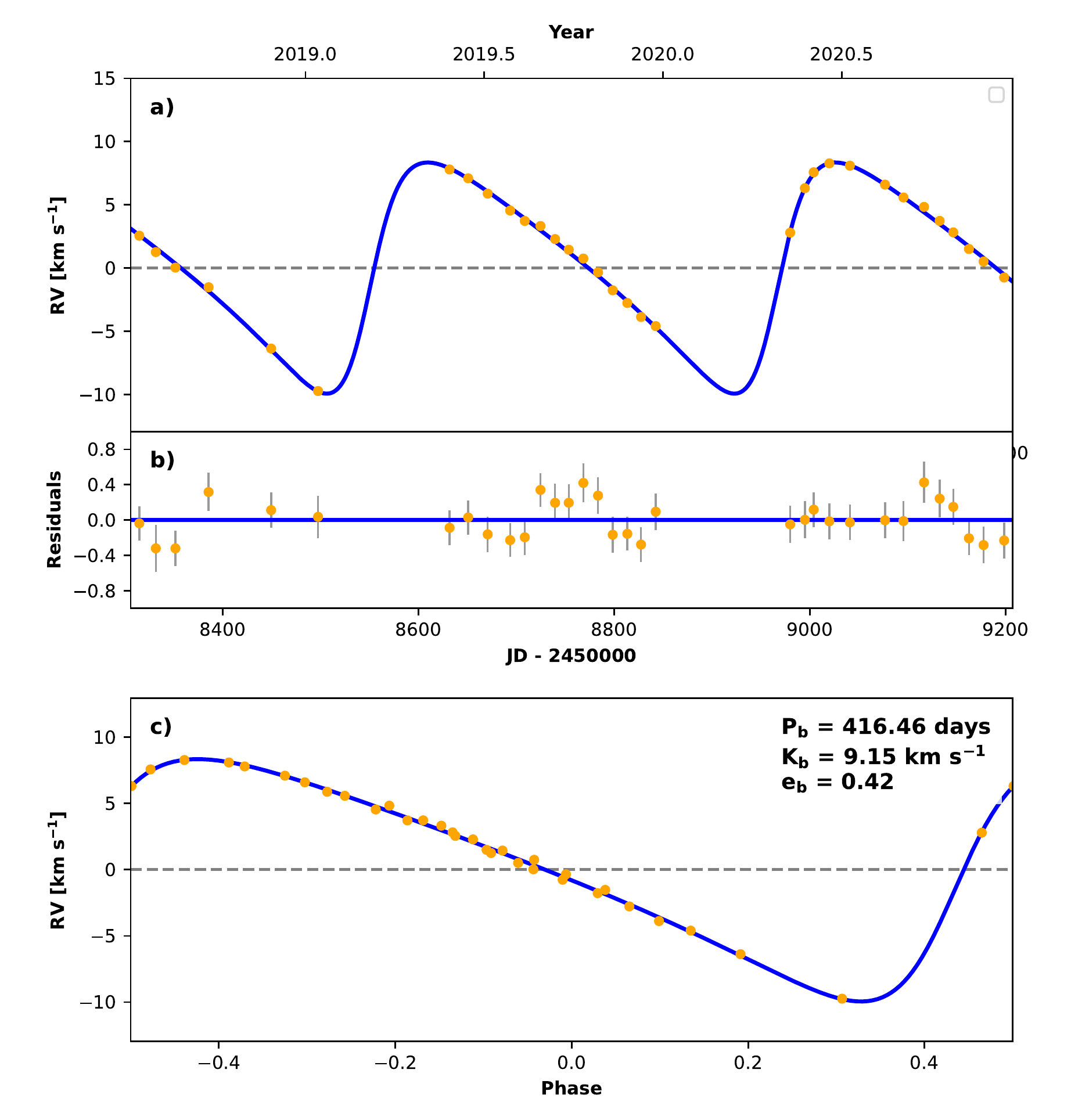}}
        \caption{RV curve of HD~217427: a) the complete measurements (circles) and RadVel fit (solid line),
        b) the residuals and c) the phase-folded RV curve.\label{fig:HD217427}}
\end{figure*}
The last star of our sample is HD~217427 and the RV curve is shown in Fig.~\ref{fig:HD217427}. The lower part
of the RV curve is not well covered. 

HD~217427 has an entry in the wide binary catalogue of \citet{tian20} 
with an angular separation of about 144~arcseconds to its companion star. 
This orbit is at a very far distance from the secondary orbit found in this work, 
so that this system is probably a hierarchical triple star system.

%
%
\section{Stellar parameters}\label{sec:param}

We determined all stellar parameters of the primary stars by analyzing the intermediate resolution
spectra ($R\approx 20,000$) obtained with the TIGRE telescope. 
By locating the stars in the
Hertzsprung-Russell diagram (HRD)
and comparing the positions with stellar evolution tracks, we were also able
to determine the masses and ages of the systems. 

\subsection{Distances, extinction and absolute magnitudes in $V$}

\begin{table*}[t]
\centering
\caption{Observational properties of five spectroscopic binary stars. 
The extinction ($E(B-V)$) was determined with the 3D dust map in order to obtain the absolute magnitudes $M_V$.}
\tabcolsep=0pt
\begin{tabular*}{500pt}{@{\extracolsep\fill}lcccccccc@{\extracolsep\fill}}
\toprule
\textbf{Star} & \textbf{Spectral type}  & $m_V$ & \textbf{parallax (mas)} & \textbf{distance (pc)} & $E(B-V)$  & $M_V$ \\
\midrule
HD 18665  & K2       & $7.25\pm0.01$ & $2.713\pm0.126$  & $373.5\pm13.1$ & $0.118\pm0.02$ & $-0.98\pm0.1$  \\  
HD 27131  & K0III+A3 & $6.94\pm0.01$ & $4.402\pm0.047$  & $228.9\pm0.8$  & $0.22\pm0.03$  & $-0.54\pm0.09$  \\  
HD 171852 & G8III/IV & $6.61\pm0.01$ & $10.897\pm0.085$ & $92.1\pm0.4$   & $0.0\pm0.01$   & $ 1.79\pm0.03$ \\   
HD 215550 & G5       & $7.20\pm0.01$ & $3.078\pm0.093$  & $328.8\pm6.2$  & $0.065\pm0.02$ & $-0.59\pm0.08$ \\
HD 217427 & K2       & $6.70\pm0.01$ & $3.764\pm0.094$  & $268.2\pm4.2$  & $0.125\pm0.02$ & $-0.83\pm0.07$ \\ 
\bottomrule
\end{tabular*}
\label{starlist}
\end{table*}

We list the observational properties like spectral type and
apparent magnitude in the $V$-band ($m_V$) as given in the SIMBAD database in Table~\ref{starlist}.
The parallaxes in milliarcseconds (and distances in parsecs) were taken from the Gaia~EDR3 archive,
and we applied the parallax zero-point correction \citep{gaiaedr32} using the published Python code.
There exist a possibly large systematic effect on the Gaia EDR3
parallaxes because of the binarity of the objects as was already discussed in Paper~I.
All stars are giants (small $M_V$), and some are located at distances, 
where reddening will affect the observed photo\-metry.
We estimated the reddening using the 3D dust mapping \citep{green19}(\url{http://argonaut.skymaps.info/}) and
obtained the color excess $E(B-V)$ using the dust map at the given positions and Gaia EDR3 distances.
We assumed an extinction law of $R(V)=3.1$ to obtain the correction for the $V$-band ($A_V$) magnitude. 
The Gaia EDR3 distances with the correction of extinction were then used to determine the absolute magnitudes in the $V$-band ($M_V$).
To determine the bolometric magnitudes and luminosities, it was necessary to apply bolometric corrections (BC),
for which exact stellar parameters, especially the effective temperature $T_\mathrm{eff}$, were required. 
Thus, a detailed analysis of the observed spectra was necessary before we could estimate
the BCs and calculate the luminosities of the five stars.

%
\subsection{Spectral analysis}

We obtained intermediate resolution spectra with a S/N that was
sufficient to determine the individual RVs.
The mean S/N of a spectrum in the red channel was about 30, while in the blue channel, depending on
the spectral type, the S/N had values between 5 and 10. 
To obtain one spectrum with a high S/N, we combined 
ten spectra with the highest S/N. 
Before, each individual spectrum was corrected for the RV.

For the determination of the stellar parameters, we used the
spectral analysis toolkit iSpec \citep{Blanco2014} in its
Python 3 version (v2020.10.01) \citep{Blanco2019}.
It allowed us to compare observed spectra with synthetic spectra determining
the selected free parameters by a $\chi^2$ method. 

We used the grid of MARCS atmosphere models \citep{Gustafsson2008} included in iSpec.
For the reference solar abundances, we used the values published in \citet{Grevesse2007}.
The synthetic spectra were calculated using the radiative transfer code {\it turbospectrum} \citep{turbo1,turbo2},
because it used spherical symmetry, which is more accurate for giant stars.
The list of atomic lines was taken from the Vienna Atomic Line Data Base (VALD) \citep{vald1,vald2}.

We have developed different methods for the determination of stellar parameters with iSpec. 
One method was used in a previous work with exoplanet hosting main sequence stars \citep{flortorres21}, and
a different method was developed as an attempt to quickly determine stellar parameters \citep{ispec21}.
In this work, we applied the improved method established in \citet{rosas21} that works very
well with giant stars.

Because we did not have any known initial values (no SIMBAD entries), 
we used an estimation of the initial values with a comparison to precalculated grid spectra.
In a first step, the improved method found the best continuum fit using individually determined continuum regions
for each spectrum based on a precalculated synthetic spectrum using the estimated initial values.
The method then used a defined set of spectral lines for the comparison between observed and theoretical spectra.
In that way, it determined the complete set of stellar parameters, which were the effective
temperature $T_\mathrm{eff}$, surface gravity $\log{g}$, metallicity [M/H], alpha enhancement [$\alpha$/Fe], 
and the micro turbulence $v_\mathrm{mic}$, macro turbulence $v_\mathrm{mac}$, and rotational velocities $v_\mathrm{rot}\sin{i}$.
We then varied the initial values for $T_\mathrm{eff}$, $\log{g}$, and [M/H] and calculated a total of 27 possible fits. 
The final fit was the one with the lowest root mean square.
Further details of the method and thorough tests were presented in \citet{rosas21}.

\begin{table*}[t]%
\centering
\caption{Basic stellar parameters of the five spectroscopic binaries determined with spectral analysis using iSpec.}
\tabcolsep=0pt%
\begin{tabular*}{500pt}{@{\extracolsep\fill}lccccccc@{\extracolsep\fill}}
\toprule
\textbf{Star} & $T_\mathrm{eff}$~[K]  & $\log{g}$ & $[M/H]$ & $[\alpha/Fe]$ & $v_\mathrm{mic}$~[km/s] &
$v_\mathrm{mac}$~[km/s] & $v_\mathrm{rot}\sin{i}$~[km/s] \\
\midrule
HD 18665  & $4253\pm 23$ & $1.81\pm0.04$ & $-0.15\pm0.02$ & $0.03\pm0.03$  & $1.93\pm0.03$ & $4.19\pm0.54$ & $0.45\pm1.63$\\
HD 27131  & $5047\pm 69$ & $2.63\pm0.14$ & $-0.31\pm0.06$ & $-0.05\pm0.06$ & $0.80\pm0.06$ & $2.40\pm2.80$ & $6.31\pm0.84$\\
HD 171852 & $5096\pm 40$ & $3.17\pm0.10$ & $-0.01\pm0.03$ & $-0.03\pm0.04$ & $1.36\pm0.05$ & $3.07\pm0.82$ & $0.00\pm0.00$\\
HD 215550 & $5063\pm 47$ & $2.76\pm0.12$ & $0.02\pm0.04$  & $-0.03\pm0.05$ & $1.33\pm0.05$ & $3.15\pm0.84$ & $0.00\pm0.00$\\
HD 217427 & $4603\pm 19$ & $2.49\pm0.05$ & $0.08\pm0.02$  & $-0.06\pm0.03$ & $1.38\pm0.03$ & $0.00\pm0.74$ & $0.00\pm0.00$\\
\bottomrule
\end{tabular*}
\label{tab:param}
\end{table*}

We present the final stellar parameters in Table \ref{tab:param}.
The effective temperatures range from about 4200 to 5100~K. The surface gravities are consistent
with the fact that all stars are giants. HD~27131 has a low metallicity of [M/H]$=-0.31$.
It is worth to state that the determination of the rotational velocity ($v_\mathrm{rot}\sin{i}$)
has improved with the new method. Some rotational velocities in Paper~I were too large.
The rotational velocities represent now expected values for giant stars, but 
it was not possible to determine rotational and also macro turbulence velocities that
have values of less than 3~km~s$^{-1}$ 
because of the intermediate resolution and the resulting instrumental profile of the HEROS spectrograph \citep{ispec21}.
For HD~27131, we found that $v_\mathrm{rot}\sin{i}=6.3$~km~s$^{-1}$.
The rotational velocities include a factor of $\sin{i}$ because of the inclination of the stellar rotation axis,
which is likely to be close to $i=90^\circ$, because we observed spectroscopic binaries. 
Of course, we assumed that the 
inclination of the rotation axis of the primary stars and of the orbits around the companion stars are similar.
Below, we will give an estimate of the orbital inclination $i$ of the binary system orbits.

\subsection{Masses and ages}

\begin{table*}[t]%
\centering
\caption{The masses and ages of the five primary stars of the spectroscopic binaries determined with fitting stellar evolution tracks to the positions in the HRD.}
\tabcolsep=0pt%
\begin{tabular*}{500pt}{@{\extracolsep\fill}lccccccc@{\extracolsep\fill}}
\toprule
\textbf{Star} & $M_V$ & BC & $M_\mathrm{bol}$ & $\log{T_\mathrm{eff}} $ & $\log{L/L_\odot} $ & \textbf{mass~$(M_\odot)$}  & \textbf{age~(Myr)} \\
\midrule
\rule{0pt}{10pt}HD 18665  & $-0.98\pm0.10$  & $-0.59 \pm 0.02$ & $-1.56 \pm 0.10$ & $3.639\pm 0.002$ & $2.55\pm0.04$ & $1.90\pm 0.05$ & $1357^{+94}_{-98}$ \\
\rule{0pt}{12pt}HD 27131  & $-0.54\pm0.09$  & $-0.23 \pm 0.02$ & $-0.77 \pm 0.10$ & $3.703\pm 0.006$ & $2.23\pm0.04$ & $2.96\pm 0.18$ & $ 401^{+124}_{-84}$ \\
\rule{0pt}{12pt}HD 171852 & $ 1.79\pm0.03$  & $-0.20 \pm 0.02$ & $1.59  \pm 0.04$ & $3.707\pm 0.003$ & $1.29\pm0.02$ & $1.95\pm 0.05$ & $1303^{+104}_{-94}$ \\
\rule{0pt}{12pt}HD 215550 & $-0.59\pm0.08$  & $-0.21 \pm 0.02$ & $-0.80 \pm 0.08$ & $3.704\pm 0.004$ & $2.24\pm0.03$ & $3.40\pm 0.05$ & $ 279^{+12}_{-11}$ \\
\rule{0pt}{12pt}HD 217427 & $-0.83\pm0.07$  & $-0.42 \pm 0.02$ & $-1.25 \pm 0.07$ & $3.663\pm 0.002$ & $2.42\pm0.03$ & $3.00\pm 0.20$ & $ 394^{+199}_{-59}$ \\
\bottomrule
\end{tabular*}
\label{tab:massage}
\end{table*}
Using the BCs calculated with the published code presented in \citet{casagrande14,casagrande18},
we obtained the absolute bolometric magnitudes $M_\mathrm{bol}$ of the stars. 
The script determined the BCs based on the three stellar parameters $T_\mathrm{eff}, \log{g}$, and [M/H].
We calculated the luminosity $L$ (in solar luminosities $L_\odot$) and present the results in Table~\ref{tab:massage}.
Having the luminosities and effective temperatures (Table~\ref{tab:param}) of the primary stars,
we could place them in the HRD.
We used stellar evolution tracks that we computed with the Cambridge (UK) Eggleton code in its updated version \citep{pols97,pols98}.
The Eggleton code overshooting parameter, and in consequence the blue loop luminosity for any given 
intermediate mass, has been accurately tested with giants of zeta Aurigae systems of well known mass and luminosity \citep{schroder97}.
The results are presented in Table~\ref{tab:massage}.
The masses ranged from 1.9 to 3.4 $M_\odot$, and the ages from 280~Myr to $\approx1.4$~Gyr.
The errors of the ages are relatively large because the evolution tracks are very close together during the giant phase.
\begin{figure}[t]
    \centerline{\includegraphics[width=0.5\textwidth]{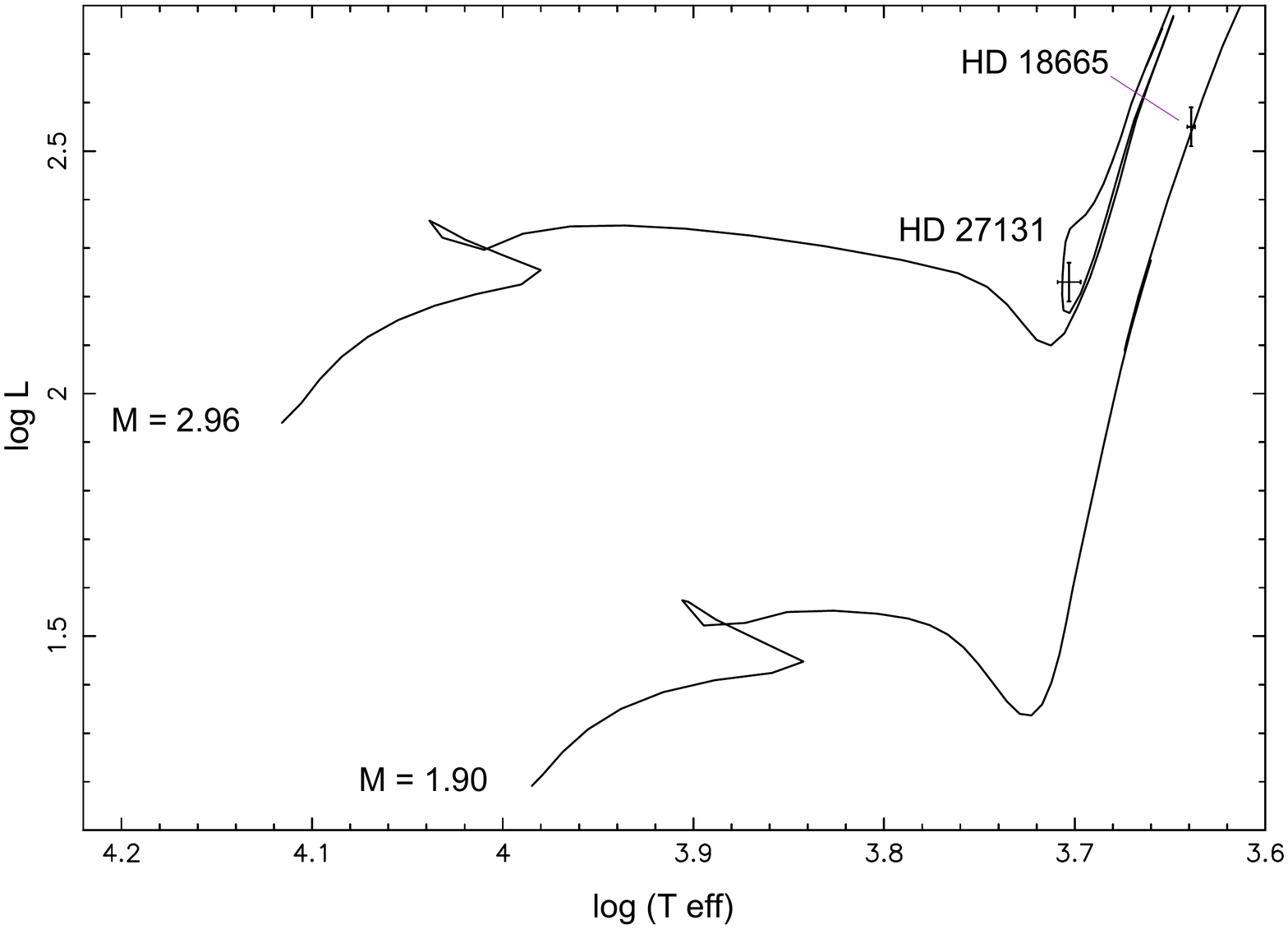}}
        \caption{Positions of HD~18665 and HD~27131 in the HRD and the best fitting stellar evolution tracks (solid lines) of the respective masses using
        tracks with $Z=0.01$.
        All stars are in the giant phase.\label{fig:tracks}}
\end{figure}
\begin{figure}[t]
    \centerline{\includegraphics[width=0.5\textwidth]{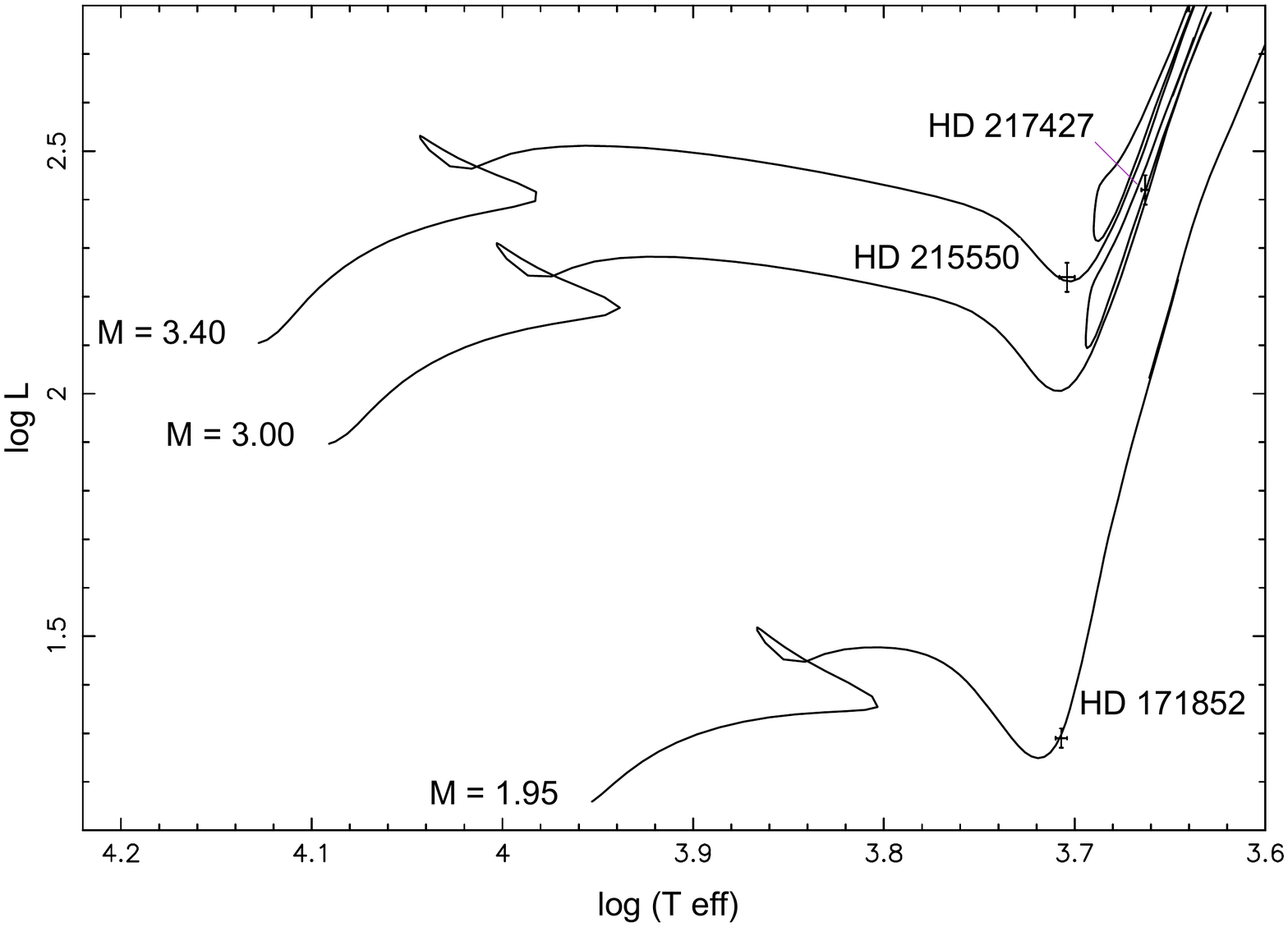}}
        \caption{Positions of HD~171852, HD~215550 and HD~217427 in the HRD and the best fitting stellar evolution tracks (solid lines) of the respective masses
        using tracks with solar metallicity ($Z=0.02$).
        All stars are in the giant phase.\label{fig:tracks2}}
\end{figure}
In Fig.~\ref{fig:tracks}, 
we show the positions of HD~18665 and HD~27131 in the theoretical HRD with $Z=0.01$ stellar evolution tracks.
The HRD positions of the other three stars (HD~171852, HD~215550, HD~217427) are presented in Fig.~\ref{fig:tracks2},
where the evolution tracks were calculated assuming solar metallicity ($Z=0.02$).
Not all stars were classified as giant stars in SIMBAD.
Please note that the errors in Table~\ref{tab:massage} are so small thanks to the stars being on the $L$-sensitive part of their blue loops.

As a consistency check, we calculated the parallax-related surface gravities of all stars as described in \citet{rosas21}.
Knowing the luminosity $L$, the mass $M_1$, and the distance from Gaia EDR3, we calculated the surface gravity using
basic relations between the stellar parameters. We found that the two
independently determined surface gravities are very similar. The largest difference
was found for HD~217427 with $\Delta \log{g}\approx0.4$, 
but this comes probably from the uncertain interstellar extinction $E(B-V)$ of that star.

\subsection{Secondary star masses and orbital inclination $i$}

\begin{table*}[t]%
\centering
\caption{The mass function ($f$) of the five spectroscopic binary systems, which we used to
determine the minimum mass of the secondary stars ($m_2$). The inclinations $i$ were determined.}
\tabcolsep=0pt%
\begin{tabular*}{500pt}{@{\extracolsep\fill}lcccccccccc@{\extracolsep\fill}}
\toprule
\textbf{Star} & $f$ ($M_\odot$) & $M_1$ ($M_\odot$) & $m_2=M_2\sin{i}$ ($M_\odot$) & $m_2^\dagger\sqrt{r}$ ($M_\mathrm{Jup}$) & $\gamma$ & $i$  & $M_2$ ($M_\odot$) \\
\midrule
\rule{0pt}{10pt}HD 18665 &  $0.0181\pm0.0006$  & $1.90\pm0.05$ & $0.41\pm0.01$  & $194^{+64}_{-40}$ & $0.204$ & $45.3^\circ ~^{+11.8^\circ}_{-6.2^\circ}$ & $0.58\pm0.1$ \\  
\rule{0pt}{12pt}HD 27131 &  $0.433 \pm0.015$   & $2.96\pm0.18$ & $1.72\pm0.07$  & $83^{+29}_{-19}$  & $0.216$ & $79.9^\circ ~^{+2.4^\circ}_{-3.3^\circ}$  & $1.75\pm0.07$ \\
\rule{0pt}{12pt}HD 171852 & $0.0933\pm0.0009$  & $1.95\pm0.05$ & $0.74\pm0.01$  & $22^{+8}_{-6}$    & $0.043$ & $69.0^\circ ~^{+5.9^\circ}_{-6.0^\circ}$  & $0.79\pm0.04$ \\
\rule{0pt}{12pt}HD 215550 & $0.2226\pm0.0011$  & $3.40\pm0.05$ & $1.45\pm0.02$  & $98^{+32}_{-19}$  & $0.207$ & $77.0^\circ ~^{+2.7^\circ}_{-3.8^\circ}$  & $1.49\pm0.05$ \\
\rule{0pt}{12pt}HD 217427 & $0.0247\pm0.0010$  & $3.00\pm0.20$ & $0.61\pm0.03$  & $12^{+30}_{-19}$  & $0.187$ & $87.0^\circ ~^{+3.0^\circ}_{-4.9^\circ}$  & $0.61\pm0.03$ \\
\bottomrule
\end{tabular*}
\label{tab:masses}
\end{table*}

As explained in more detail in Paper~I, one can use the Kepler's laws 
to derive a formula that connects the masses of the
two components of a binary system with three observable orbital parameters,
which are the orbital period $P$, semi-amplitude $K$, and the eccentricity $e$.
This so called mass function $f$ is calculated by the formula
\begin{equation}
\label{eq1}
f=\frac{M_2^3\sin^3{i}}{(M_1+M_2)^2}=\frac{P K^3}{2\pi G}(1-e^2)^{3/2},
\end{equation}
where $G$ represents the gravitational constant. 
The mass function $f$ allows to calculate both masses from one spectroscopic orbit
only, if both inclination $i$ and the mass ratio are known. This is only the case for eclipsing SB2
systems (i.e., $\sin{i}$ close to 1). Alternatively, $\sin{i}$ can be obtained from an astrometric orbit,
where available. However, in the cases of ordinary SB1 systems, as treated by this paper, 
the strategy can only be to obtain the masses, and hence the mass ratio, 
from the photometric parameters and parallax of the system. 
In turn, the mass function then allows to estimate $\sin{i}$.
Because we determined the masses of the primary
stars ($M_1$), we could calculate the minimal masses of the secondary stars ($m_2=M_2\sin{i}$) solving Eq.~\ref{eq1}.
We present the values for the mass function $f$ and the minimal secondary masses ($m_2$) in Table~\ref{tab:masses}.


Our five binary stars were also identified as binaries by \citet{kervella19} in an 
analysis of the proper motion anomaly of Hipparcos stars using Gaia DR2 data.
They also determined the masses of the secondary stars normalized for a 1~AU circular orbit $m_2^\dagger$,
which was related to the minimal mass of the secondary by the relation $m_2^\dagger\sqrt{r}$, 
where $r$ was the orbital distance (in AU) from the primary, unknown for them.

We could calculate the minimal masses of \citet{kervella19} using the orbital distance $r$ from our orbital parameters using the Kepler's laws. 
The orbital distance of the primary is given by
\begin{equation}
r = a_1 =\left(\frac{G}{4\pi^2}\frac{M_2^3}{M_\mathrm{tot}^2}P^2\right)^{1/3}.
\end{equation}
We could then calculate the minimal mass of the secondary assuming a circular orbit.
The orbit of the binary is projected in the sky, 
and the variation in the proper motion has its maximum for an inclination of $i=0^\circ$.
This is the complete opposite projection
than for the orbit determined with RV measurements, where the maximum effect is found at
an inclination of $i=90^\circ$. 
The \citet{kervella19} minimal masses are related to the actual secondary mass $M_2$ by $\cos{i}$, 
while the minimal masses from the RV curves are related to the actual mass by $\sin{i}$.
Combining these two minimal masses, we determined the orbital inclination $i$ using the equation
\begin{equation}
i=\arctan{\frac{m_2}{m_2^\dagger\sqrt{r}}}.
\end{equation}
In addition, we calculated the actual mass of the secondary with the equation $M_2=m_2/\sin{i}$. 

For orbital periods that are smaller or close to the observing window of Gaia DR2 ($\delta t=668$~days), 
the observed effect of a proper motion anomaly will decrease because of a smearing effect of the signal. 
\citet{kervella19} demonstrated the impact of this effect in their Figure~2. 
The graph showed us that binary systems with periods presented in this work 
are affected by this smearing effect with a factor of roughly $\gamma\approx 0.2$.
Because we know the exact period of each binary system, we could calculate the sensitivity factor $\gamma$
using Equation 13 of \citet{kervella19}. This factor increased the actual signal and increased 
their normalized minimal secondary masses $m_2^\dagger$.

The factor $\gamma$ and the results for the inclination $i$ and the secondary mass $M_2$ are presented in Table~\ref{tab:masses}.
All orbital inclinations $i$ are closer to $90^\circ$ than to $0^\circ$. The lowest inclination
has the star HD~18665 ($i=45.3^\circ$).
Because of the large inclinations, the masses of the secondary stars $M_2$ 
are very similar to the minimal masses $m_2$ determined before. 
Even for HD~18665 ($m_2=0.41\;M_\odot$) the secondary mass is still less than solar ($M_2=0.58\;M_\odot$).

It is important to mention that the orbital inclinations determined 
with this method are very sensitive to the values of \citet{kervella19}. 
The low sensitivity factor $\gamma$ increased this even further.
Thus, the inclinations presented here may not be very reliable.
There is even the possibility that they measured the effect of a different companion star and further
member of the system or a combination effect of several stars in a multiple stellar system.

%
%
\section{Summary}\label{sec:con}

In the second publication of our series about bright spectroscopic binaries,
we determined the stellar and orbital parameters of five binary systems
with orbital periods of about $P\lesssim500$~days. Using intermediate resolution spectra, 
we determined the stellar parameters of the five stars 
with an improved method for iSpec.
With Gaia~EDR3 parallaxes considering interstellar extinction and BCs,
we determined the masses and ages by a comparison of the HRD positions with stellar evolution tracks
calculated with the Eggleton code.
Finally, an estimation of the masses of the respective secondary stars were presented.
Using the results of an analysis of proper motions anomaly of \citet{kervella19},
we estimated the orbital inclinations of the binary systems and could determine the
absolute secondary star masses.
We found that three of the primary stars were not classified as giants in the SIMBAD database,
and no star as spectroscopic binary.

\section*{Acknowledgments}

This research has been made possible by the CONACyT-DFG bilateral grant No. 278156.
We thank the University of Guanajuato for the grants for the projects 036/2021 and 105/2021
of the {\it Convocatoria Institucional de Investigaci\'on Cient\'ifica 2021}.
This work has made use of data from the European Space Agency (ESA) mission
{\it Gaia} (\url{https://www.cosmos.esa.int/gaia}), processed by the {\it Gaia}
Data Processing and Analysis Consortium (DPAC,
\url{https://www.cosmos.esa.int/web/gaia/dpac/consortium}). Funding for the DPAC
has been provided by national institutions, in particular the institutions
participating in the {\it Gaia} Multilateral Agreement.
This work has made use of the VALD database, operated at Uppsala University, 
the Institute of Astronomy RAS in Moscow, and the University of Vienna.
This research has made use of the SIMBAD database, operated at CDS, Strasbourg, France.
This research has made use of the VizieR catalogue access tool, CDS, Strasbourg, France \citep{vizier}.

\bibliography{all}



\appendix

\section{RV measurements}
\label{Sec:RV}

\begin{table}[t]%
\centering
\caption{RV measurements of all stars (also electronically available as supporting
online material at \emph{link to Wiley page here} as well as at \url{http://www.astro.ugto.mx/~dennis/binaries/}).}
\tabcolsep=0pt%
\begin{tabular}{lcc}
\toprule
Stars & JD (days) & $RV$ (km$^{-1}$)  \\
\midrule
\textbf{HD 18665}~~ & 2458366.84840 & ~~ $-5.13 \pm 0.12$ \\
 & 2458397.78679 & ~~ $-6.06 \pm 0.13$ \\
 & 2458446.74705 & ~~ $-7.98  \pm 0.11$ \\
 & 2458466.56093 & ~~ $-8.84 \pm  0.14$ \\
 & 2458493.57980 & ~~ $-9.01 \pm  0.09$ \\
 & 2458514.57933 & ~~ $-9.27 \pm  0.09$ \\
 & 2458518.57777 & ~~ $-9.12 \pm  0.10$ \\
 & 2458692.97971 & ~~ $0.18  \pm  0.12$ \\
 & 2458725.83523 & ~~ $5.94 \pm  0.11$ \\
 & 2458751.78164 & ~~ $4.30 \pm  0.12$ \\
 & 2458771.67443 & ~~ $1.99 \pm  0.13$ \\
 & 2458794.69108 & ~~ $-0.22 \pm  0.13$ \\
 & 2458814.61678 & ~~ $-2.45 \pm  0.11$ \\
 & 2458832.57071 & ~~ $-3.70 \pm  0.11$ \\
 & 2458854.58975 & ~~ $-4.45 \pm  0.11$ \\
 & 2458875.65306 & ~~ $-5.61 \pm  0.11$ \\
 & 2459045.97341 & ~~ $-10.29 \pm  0.10$ \\
 & 2459085.84083 & ~~ $-10.10 \pm  0.16$ \\
 & 2459118.81388 & ~~ $-9.92 \pm  0.11$ \\
 & 2459141.78061 & ~~ $-8.61 \pm  0.13$ \\
 & 2459164.72791 & ~~ $-6.88 \pm  0.11$ \\
 & 2459195.63323 & ~~ $-1.12 \pm  0.11$ \\
 & 2459248.58735 & ~~ $4.88 \pm  0.10$ \\
\textbf{HD 27131}~~ & 2458388.99545 & ~~ $1.72 \pm 0.16$ \\
 & 2458446.78764 & ~~ $11.52 \pm 0.16$ \\
 & 2458461.73016 & ~~ $16.84 \pm 0.13$ \\
 & 2458472.75425 & ~~ $20.49 \pm 0.13$ \\
 & 2458491.64267 & ~~ $27.27 \pm 0.18$ \\
 & 2458504.65623 & ~~ $31.84 \pm 0.08$ \\
 & 2458519.58549 & ~~ $36.51 \pm 0.12$ \\
 & 2458539.61070 & ~~ $41.31 \pm 0.20$ \\
 & 2458716.98552 & ~~ $18.54 \pm 0.14$ \\
 & 2458728.98998 & ~~ $15.85 \pm 0.14$ \\
 & 2458743.99337 & ~~ $12.63 \pm 0.15$ \\
 & 2458760.84030 & ~~ $9.871 \pm 0.17$ \\
 & 2458770.87912 & ~~ $8.19 \pm 0.18$ \\
 & 2458784.83426 & ~~ $6.08 \pm 0.15$ \\
 & 2458796.76591 & ~~ $4.78 \pm 0.26$ \\
 & 2458807.76734 & ~~ $3.55 \pm 0.15$ \\
 & 2458817.66481 & ~~ $3.13 \pm 0.10$ \\
 & 2458827.65304 & ~~ $2.78 \pm 0.15$ \\
 & 2458835.56184 & ~~ $2.95 \pm 0.13$ \\
 & 2458845.69483 & ~~ $3.03 \pm 0.22$ \\
 & 2458856.64204 & ~~ $4.19 \pm 0.13$ \\
\bottomrule
\end{tabular}
\label{tab:rv_1}
\end{table}

\setcounter{table}{0}
\begin{table}[t]%
\centering
\caption{continued.}
\tabcolsep=0pt%
\begin{tabular}{lcc}
\toprule
Stars & JD (days) & $RV$ (km$^{-1}$)  \\
\midrule
 & 2458866.66379 & ~~ $5.39 \pm 0.14$ \\
 & 2458877.58959 & ~~ $6.98 \pm 0.11$ \\
 & 2459076.98692 & ~~ $41.53 \pm 0.19$ \\
 & 2459120.83843 & ~~ $32.57 \pm 0.16$ \\
 & 2459134.81739 & ~~ $29.34 \pm 0.20$ \\
 & 2459144.83482 & ~~ $26.94 \pm 0.21$ \\
 & 2459158.73090 & ~~ $24.09 \pm 0.40$ \\
 & 2459172.77316 & ~~ $20.50 \pm 0.26$ \\
 & 2459193.72306 & ~~ $16.02 \pm 0.15$ \\
 & 2459204.66835 & ~~ $14.00 \pm 0.15$ \\
 & 2459219.65747 & ~~ $11.19 \pm 0.16$ \\
 & 2459232.69133 & ~~ $9.22 \pm 0.15$ \\
 & 2459255.58633 & ~~ $6.37 \pm 0.16$ \\
\textbf{HD 171852}~~ & 2458250.76875 & ~~ $-22.75 \pm 0.15$ \\
 & 2458262.97179 & ~~ $-24.91 \pm 0.08$ \\
 & 2458272.87976 & ~~ $-26.44 \pm 0.15$ \\
 & 2458276.96559 & ~~ $-27.10 \pm 0.13$ \\
 & 2458301.72960 & ~~ $-30.72 \pm 0.11$ \\
 & 2458315.79259 & ~~ $-32.69 \pm 0.09$ \\
 & 2458324.78002 & ~~ $-34.02 \pm 0.11$ \\
 & 2458331.82748 & ~~ $-34.67 \pm 0.20$ \\
 & 2458353.67498 & ~~ $-37.83 \pm 0.23$ \\
 & 2458385.59252 & ~~ $-40.67 \pm 0.17$ \\
 & 2458404.58398 & ~~ $-39.69 \pm 0.68$ \\
 & 2458530.02143 & ~~ $-13.22 \pm 0.11$ \\
 & 2458547.01505 & ~~ $-14.84 \pm 0.10$ \\
 & 2458563.00563 & ~~ $-17.12 \pm 0.08$ \\
 & 2458579.97171 & ~~ $-19.61 \pm 0.13$ \\
 & 2458598.89813 & ~~ $-23.03 \pm 0.16$ \\
 & 2458608.92355 & ~~ $-24.35 \pm 0.26$ \\
 & 2458623.78951 & ~~ $-26.71 \pm 0.10$ \\
 & 2458638.86077 & ~~ $-29.11 \pm 0.11$ \\
 & 2458660.73142 & ~~ $-32.48 \pm 0.09$ \\
 & 2458675.74263 & ~~ $-34.35 \pm 0.07$ \\
 & 2458695.65868 & ~~ $-36.99 \pm 0.09$ \\
 & 2458712.69842 & ~~ $-39.11 \pm 0.10$ \\
 & 2458730.65281 & ~~ $-40.59 \pm 0.16$ \\
 & 2458748.56696 & ~~ $-41.10 \pm 0.12$ \\
 & 2458762.55909 & ~~ $-40.51 \pm 0.14$ \\
 & 2458772.56066 & ~~ $-39.37 \pm 0.11$ \\
 & 2458796.55364 & ~~ $-32.04 \pm 0.18$ \\
 & 2458898.02129 & ~~ $-15.09 \pm 0.08$ \\
 & 2458918.00766 & ~~ $-18.12 \pm 0.09$ \\
 & 2458934.00207 & ~~ $-20.71 \pm 0.13$ \\
 & 2458949.99080 & ~~ $-23.21 \pm 0.09$ \\
 & 2458963.98214 & ~~ $-25.47 \pm 0.09$ \\
 & 2458981.97501 & ~~ $-28.24 \pm 0.13$ \\
\bottomrule
\end{tabular}
\label{tab:rv_2}
\end{table}
%
\setcounter{table}{0}

\begin{table}[t]%
\centering
\caption{continued.}
\tabcolsep=0pt%
\begin{tabular}{lcc}
\toprule
Stars & JD (days) & $RV$ (km$^{-1}$)  \\
\midrule

 & 2458995.86851 & ~~ $-30.06 \pm 0.12$ \\ 
 & 2459015.82626 & ~~ $-33.19 \pm 0.08$ \\
 & 2459042.75698 & ~~ $-36.83 \pm 0.08$ \\
 & 2459076.64807 & ~~ $-40.35 \pm 0.11$ \\
 & 2459111.60202 & ~~ $-40.71 \pm 0.15$ \\
 & 2459127.56708 & ~~ $-37.77 \pm 0.10$ \\
 & 2459145.56472 & ~~ $-32.40 \pm 0.13$ \\
 & 2459166.54059 & ~~ $-23.34 \pm 0.12$ \\
 & 2459264.02113 & ~~ $-17.68 \pm 0.10$ \\
 & 2459293.00521 & ~~ $-22.52 \pm 0.13$ \\
 & 2459310.97814 & ~~ $-25.08 \pm 0.14$ \\
 & 2459324.98516 & ~~ $-27.43 \pm 0.09$ \\
 
\textbf{HD 215550}~~ & 2458314.95635 & ~~ $-15.59 \pm 0.13$ \\
 & 2458331.81999 & ~~ $-18.02 \pm 0.19$ \\
 & 2458349.82508 & ~~ $-18.20 \pm 0.15$ \\
 & 2458376.77598 & ~~ $-15.34 \pm 0.15$ \\
 & 2458387.69197 & ~~ $-14.30 \pm 0.14$ \\
 & 2458401.70564 & ~~ $-13.05 \pm 0.10$ \\
 & 2458449.59715 & ~~ $-6.94 \pm 0.11$ \\
 & 2458464.61000 & ~~ $-4.96 \pm 0.23$ \\
 & 2458476.54501 & ~~ $-3.26 \pm 0.11$ \\
 & 2458494.56731 & ~~ $-0.91 \pm 0.15$ \\
 & 2458608.95536 & ~~ $14.60 \pm 0.14$ \\
 & 2458623.96844 & ~~ $16.01 \pm 0.12$ \\
 & 2458648.93751 & ~~ $18.35 \pm 0.11$ \\
 & 2458668.91267 & ~~ $17.92 \pm 0.10$ \\
 & 2458690.82169 & ~~ $13.50 \pm 0.22$ \\
 & 2458707.78405 & ~~ $5.49 \pm 0.16$ \\
 & 2458720.76589 & ~~ $-2.28 \pm 0.16$ \\
 & 2458730.71462 & ~~ $-7.740 \pm 0.10$ \\
 & 2458741.74124 & ~~ $-12.31 \pm 0.17$ \\
 & 2458751.63339 & ~~ $-15.47 \pm 0.12$ \\
 & 2458761.66417 & ~~ $-17.14 \pm 0.12$ \\
 & 2458770.59163 & ~~ $-17.75 \pm 2 0.10$ \\
 & 2458783.60943 & ~~ $-18.33 \pm 0.16$ \\
 & 2458793.58315 & ~~ $-17.69 \pm 0.15$ \\
 & 2458813.56829 & ~~ $-16.06 \pm 0.10$ \\
 & 2458820.60321 & ~~ $-15.28 \pm 0.13$ \\
 & 2458830.56476 & ~~ $-14.13 \pm 0.12$ \\
 & 2458841.54270 & ~~ $-12.72 \pm 0.09$ \\
 & 2458977.97469 & ~~ $6.09 \pm 0.34$ \\
 & 2458987.96966 & ~~ $6.34 \pm 0.14$ \\
 & 2458995.96976 & ~~ $7.48 \pm 0.14$ \\
 & 2459000.97078 & ~~ $8.31 \pm 0.13$ \\
 & 2459012.96938 & ~~ $9.86 \pm 0.14$ \\
 & 2459033.94508 & ~~ $12.89 \pm 0.17$ \\
 & 2459045.85052 & ~~ $14.16 \pm 0.18$ \\

\bottomrule
\end{tabular}
\label{tab:rv_3}
\end{table}
%
\setcounter{table}{0}

\begin{table}[t]%
\centering
\caption{continued.}
\tabcolsep=0pt%
\begin{tabular}{lcc}
\toprule
Stars & JD (days) & $RV$ (km$^{-1}$)  \\
\midrule
 & 2459076.88206 & ~~ $17.04 \pm 0.13$ \\
 & 2459088.75662 & ~~ $18.01 \pm 0.17$ \\
 & 2459106.71727 & ~~ $17.76 \pm  0.14$ \\
 & 2459117.69385 & ~~ $16.33 \pm 0.16$ \\
 & 2459127.67240 & ~~ $13.46 \pm 0.31$ \\
 & 2459141.61819 & ~~ $6.90 \pm 0.10$ \\
 & 2459151.60453 & ~~ $1.29 \pm 0.11$ \\
 & 2459162.60925 & ~~ $-5.34 \pm 0.14$ \\
 & 2459172.57873 & ~~ $-10.13 \pm 0.25$ \\
 & 2459182.54953 & ~~ $-14.22 \pm 0.11$ \\
 & 2459195.54626 & ~~ $-16.99 \pm 0.14$ \\
 & 2459208.57139 & ~~ $-18.02 \pm 0.14$ \\
 & 2459328.98114 & ~~ $-6.53 \pm 0.21$ \\
\textbf{HD 217427}~~ & 2458314.95475 & ~~ $ -22.47 \pm 0.09$ \\
 & 2458331.82149 & ~~ $ -23.76 \pm 0.20$ \\
 & 2458351.80596 & ~~ $ -24.99 \pm 0.11$ \\
 & 2458385.71645 & ~~ $ -26.54 \pm 0.13$ \\
 & 2458449.59554 & ~~ $ -31.40 \pm 0.10$ \\
 & 2458497.56869 & ~~ $ -34.76 \pm 0.17$ \\
 & 2458631.95965 & ~~ $ -17.23 \pm 0.10$ \\
 & 2458650.96811 & ~~ $ -17.92 \pm 0.09$ \\
 & 2458670.91219 & ~~ $ -19.14 \pm 0.10$ \\
 & 2458693.83997 & ~~ $ -20.47 \pm 0.08$ \\
 & 2458708.79431 & ~~ $ -21.30 \pm 0.10$ \\
 & 2458724.82595 & ~~ $ -21.70 \pm 0.08$ \\
 & 2458739.74297 & ~~ $ -22.73 \pm 0.13$ \\
 & 2458753.70438 & ~~ $ -23.58 \pm 0.12$ \\
 & 2458768.63801 & ~~ $ -24.27 \pm 0.14$ \\
 & 2458783.64526 & ~~ $ -25.37 \pm 0.11$ \\
 & 2458798.61809 & ~~ $ -26.79 \pm 0.11$ \\
 & 2458813.56671 & ~~ $ -27.79 \pm 0.08$ \\
 & 2458827.58366 & ~~ $ -28.90 \pm 0.09$ \\
 & 2458842.55102 & ~~ $ -29.62 \pm 0.12$ \\
 & 2458979.97344 & ~~ $ -22.22 \pm 0.12$ \\
 & 2458994.96788 & ~~ $ -18.71 \pm 0.12$ \\
 & 2459003.97033 & ~~ $ -17.45 \pm 0.10$ \\
 & 2459019.96992 & ~~ $ -16.74 \pm 0.11$ \\
 & 2459040.97416 & ~~ $ -16.93 \pm 0.10$ \\
 & 2459076.84217 & ~~ $ -18.42 \pm 0.11$ \\
 & 2459095.72271 & ~~ $ -19.44 \pm 0.15$ \\
 & 2459116.71672 & ~~ $ -20.19 \pm 0.16$ \\
 & 2459132.69030 & ~~ $ -21.29 \pm 0.13$ \\
 & 2459146.57782 & ~~ $ -22.20 \pm 0.11$ \\
 & 2459162.61127 & ~~ $ -23.52 \pm 0.08$ \\
 & 2459177.54677 & ~~ $ -24.51 \pm 0.12$ \\
 & 2459198.54290 & ~~ $ -25.78 \pm 0.11$ \\
\bottomrule
\end{tabular}
\label{tab:rv_4}
\end{table}

We present all TIGRE RV measurements of the five binary stars in Table~ \ref{tab:rv_1}.

\end{document}